\documentstyle[twocolumn,aps,prl,overcite]{revtex} 
\input epsf
\date{\today}

\tighten
\begin{document}

\twocolumn[\hsize\textwidth\columnwidth\hsize\csname @twocolumnfalse\endcsname

%%%%%%%%%%%%%%%%%%%%%%%%%%%%%%%%%%%%%%%%%%%%%%%%%%%%%%%%%%%%%%%%%%%%%%%%%%

\title{ Resolving the Cosmological Missing Energy Problem}

\author{Greg Huey, Limin Wang, R. Dave, R. R. Caldwell, 
and Paul J. Steinhardt }

\address{Department of Physics and Astronomy\\
University of Pennsylvania\\
Philadelphia, PA 19104}

\maketitle

%%%%%%%%%%%%%%%%%%%%%%%%%%%%%%%%%%%%%%%%%%%%%%%%%%%%%%%%%%%%%%%%%%%%%%%%%%
\begin{abstract}

Some form of missing energy may account for the difference between the
observed cosmic matter density and the critical density. Two leading
candidates are a cosmological constant and quintessence (a time-varying,
inhomogenous component with negative pressure). We show that an ideal,
full-sky cosmic background anisotropy experiment may not be able to
distinguish the two, even when non-linear effects due to gravitational
lensing are included.  Due to this ambiguity,  microwave background
experiments alone may not determine the matter density or Hubble constant
very precisely.  We further show that degeneracy may remain even after
considering classical cosmological tests and measurements of large scale
structure.

\end{abstract} 
]
 
This paper looks ahead a few years to a time when highly precise,
full-sky maps of the cosmic microwave background (CMB) anisotropy become
available from satellite experiments such as the NASA Microwave
Anisotropy Probe\cite{MAP}  (MAP) and the ESA Planck
mission.\cite{Planck}  The goal is to determine if  measurements of the
anisotropy by itself or combined with other cosmological constraints can
resolve between competing models for the ``missing energy" of the
universe. The missing energy problem arises because inflationary
cosmology and  some current microwave anisotropy measurements suggest
that the universe is flat at the same time that a growing number of
observations  indicate that the matter density (baryonic and nonbaryonic)
is below the critical density  ($\Omega_m < 1$).\cite{MEP}  These two
trends can be reconciled if there is another contribution to the  energy
density of the universe besides matter. One candidate for the missing
energy is a vacuum density or cosmological constant
($\Lambda$).\cite{Weinberg,CPT,TurnerKrauss,Ostriker}   A second
candidate  is quintessence, a time-varying, spatially inhomogeneous
component with negative pressure.\cite{CDS}   Both models fit all current
observations well.\cite{Ostriker,WCOS}  

If current observational trends continue,  determining the nature of the
missing energy will emerge as one of cosmology's most important
challenges. The issue must be decided in order to  understand  the energy
composition of the universe. Also, as  shown below, ambiguity concerning
the missing energy leads to large uncertainties in  two key parameters:
$\Omega_m$ and $h$ (the Hubble constant in units of
100~km~sec$^{-1}$~Mpc$^{-1}$). In this paper, we show that, despite
extraordinary advances  in measurements of the CMB anisotropy and
large-scale structure anticipated in the near future, the missing energy
problem  and, consequently $\Omega_m$ and $h$, may remain unresolved in
some circumstances.

The key differences between quintessence and vacuum density are: (1)
quintessence has an equation-of-state $w$ (equal to the ratio of 
pressure to energy density) greater than $-1$, whereas vacuum density 
has $w$ precisely equal to $-1$;   (2)  the energy density for
quintessence varies with time whereas the vacuum density is constant; and
(3), quintessence is spatially inhomogeneous  and can cluster
gravitationally,  whereas vacuum density remains spatially uniform.  The
first two properties result in different predictions for the expansion
rate.  The third property  results in     a direct imprint of
quintessence fluctuations on the  CMB and large scale structure.

For the purposes of this investigation, we model quintessence as a cosmic
scalar field $Q$ evolving in a potential, $V(Q)$.  Depending on the form
of $V(Q)$,  the equation-of-state $w$ can be constant,  monotonically
increasing or decreasing, or oscillatory.\cite{CDS,CDSconf}  If $w$  is
time-varying, it is useful to define an average equation-of-state as
$\bar{w} \approx \int da\,\Omega_Q(a) w(a) /\int da\,\Omega_Q(a)$ where 
$a$ is the expansion scale factor. Roughly speaking, the CMB temperature
and the mass power spectra of a model with a slowly-varying $w(a)$ is
most similar to those of a constant  $w$ model with $w=\bar{w}$. We can
also define $\dot{\bar{w}}^2 \equiv \int dz\,\Omega_Q(z) [\dot{w}]^2
/\int dz\, \Omega_Q(z)$, where $\dot{w} \equiv dw/d\ln{z}$. If $w$ is
rapidly varying, $\dot{\bar{w}}^2 \gtrsim 1$, the spatial fluctuations
in  $Q$ and the variation in the cosmic expansion rate significantly
alter the  shape of the cosmic microwave anisotropy power
spectrum,\cite{CDS,CDSconf}  producing differences from  $\Lambda$ models
that are detectable in near-future satellite measurements.   

The degeneracy problem between $\Lambda$ and quintessence arises if $w$
is constant or slowly-varying ($\dot{\bar{w}}^2 \ll 1$), as occurs for  a
wide range of potentials ({\it e.g.}, quadratic or exponential) and
initial conditions.  It is well recognized that two models with identical
primordial perturbation spectra, matter content at last scattering, and 
comoving distance to the surface of last scattering  generate
statistically identical linear CMB power spectra.\cite{Sperg97,BondEtAl}
This situation is referred to as the geometric degeneracy, owing to the
identical geometrical optics of the  comoving line of sight and sound
horizon at last scattering.  In this case, we find that for $w \lesssim
-\Omega_Q/2$, the effects of quintessence on the CMB power spectrum as
will be observed by MAP ($\ell \lesssim 800$)  can be closely mimicked by
a model with $\Lambda$, provided  $\Omega_m$ and $h$ are also adjusted.  

Measurements of the CMB on smaller angular scales where non-linear
effects are important  can be used to break the degeneracy. Gravitational
lensing distortion of the primary, linear CMB anisotropy by small-scale
density inhomogeneities along the line of
sight\cite{BlanchEtAl,ColeEfstathiou,CayonEtAl,Seljak} has the capability
to discriminate\cite{Metcalf,StomporEfstathiou} between quintessence and
a cosmological constant.\cite{4by4_properties} The efficacy of this
phenomena, which smooths the peaks and troughs in the CMB spectrum on the
scales of interest, depends on the level of mass fluctuations. If the
amplitude of primordial density perturbations were anything other than
$\delta\rho/\rho \sim 10^{-5}$, this effect would be either completely
negligible or else the dominant effect in CMB anisotropy.  At the level
measured by COBE  and MAP, the lensing is a negligible effect since it
only begins to become important for $\ell \gtrsim 1000$.   However,
lensing effects are non-negligible for the Planck experiment which 
extends to $\ell \sim 1500$, or experiments at yet smaller angular 
scales.

\begin{figure}
\epsfxsize=3.3 in \epsfbox{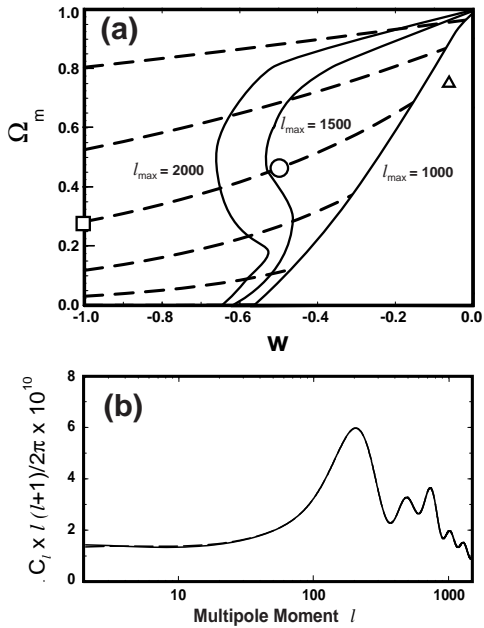} 
\caption{The CMB degeneracy problem: Each dashed curve in (a)  represents
a family of QCDM and $\Lambda$CDM models with indistinguishable CMB
anisotropy power spectra.  For an ideal, full-sky, cosmic variance
limited experiment with $\ell_{\rm max}= 1000,\,1500,\,2000$, the solid
lines mark the projection of the right-most boundary of the degeneracy
region. For example, Panel (b) shows  two overlapping  spectra for the 
$\Lambda$  (square) and quintessence  (circle) models indicated in (a).
Models beyond the solid line in (a) ({\it e.g.}, the triangle for
$\ell_{\rm max}=1000$, or the circle for $\ell_{\rm max}=2000$) are
distinguishable. }
\end{figure}

Figure 1 illustrates the degeneracy problem for CMB anisotropy
measurements.  Figure 1a shows the $\Omega_m-w$ plane of quintessence
models with a slowly varying or constant equation of state, where the
axis $w=-1$ corresponds to the case of a cosmological constant. Each
dashed curve represents a set of cosmological models with a $Q$- or
$\Lambda$-component which satisfy the conditions for the  geometric
degeneracy for CMB anisotropy power spectra.  The solid curves represent
the projection of the right-most border of the degeneracy region,  in the
full parameter space, at which models can be distinguished from
$\Lambda$CDM at the $\ge 3\sigma$ level, including the effect of lensing,
for an idealized, cosmic variance limited experiment with a given maximum
multipole moment.   For example, for fixed $\Omega_b h^2$ and  $n_s$ (the
spectral index of scalar fluctuations), a model with quintessence and
$\Omega_m=0.47$, $w=-1/2$ and $h=0.57$ (circle)  produces a nearly
identical CMB power spectrum to a $\Lambda$ model with $\Omega_m=0.29$,
$w=-1$ and $h=0.72$ (square) for $\ell_{\rm max} = 1000$. As the range of
multipole moments increases, including smaller scale CMB anisotropy, the
gravitational lensing distortion becomes more pronounced, breaking the
geometric degeneracy. Yet we see that for many quintessence models,  even
with an ideal, cosmic variance limited, full-sky  measurement of the CMB
anisotropy with multipoles $\ell \le 2000$, there  remains a degeneracy
in the $\Omega_m - w$ parameter space. We show Figure 1b to illustrate
the extent  of the degeneracy, as the two power spectra overlap almost
entirely.

The degeneracy curves can be understood theoretically. They correspond
approximately to the set of models that obey the following constraints:
(a) $\Omega_m  +  \Omega_{Q} =   A = 1$;  (b) $\Omega_m h^2  =   B$; (c)
$\Omega_b h^2   =  C$; (d)  $n_s  =  D$;   and, (e) $\ell_P  =  E$. 
Here  $A, B, C, D, E$  are  constants, and $\ell_P$ is the multipole
corresponding the position of the first acoustic (Doppler) peak.  
Constraint (a) is the flatness condition. Constraints (b)-(d) are
required in order for the Doppler peak heights to remain constant.  Along
with constraint (d), we assume that $r$, the ratio of the
tensor-to-scalar primordial power spectrum amplitudes obeys inflationary
predictions.\cite{Davis,CaldwellSteinhardt} Constraint (e) insures that
the acoustic peaks occur at the same multipole moment. The peak position
$\ell_P$ (proportional to the ratio of the conformal time since last
scattering to the  sound horizon at last scattering) depends on 
$\Omega_m h^2$, $\Omega_b h^2$, $h$ and $w$. The only way to keep
$\ell_P$ constant along the degeneracy curve as $w$ varies is to adjust
$h$, since $\Omega_m h^2$ and $\Omega_b h^2$ are constrained to be fixed
by (b) and (c). (M. White has independently noted  similar
conditions for degeneracy for  constant $w$ models.\cite{White98})   Our
results are based on full numerical codes which include the  fluctuations
in $Q$ and the gravitational lensing distortion\cite{CMBFAST}.  Our
computations  confirm that the above conditions are a good approximation
to  the degeneracy curves. When we restrict our attention to
$\ell\lesssim 1000$, angular scales on which lensing is negligible, then
if the value of $h$ for the first model  along the degeneracy curve is
changed,  the value of $h$ for the rest of the models can be adjusted so
that the geometric degeneracy remains.  The boundary of the $\ell_{\rm
max}=1000$ degeneracy region is then determined by fluctuating $Q$
effects and the large integrated Sachs-Wolfe contribution to the  CMB
anisotropy, such that models with $w \gtrsim -\Omega_Q/2$  are
distinguishable from $\Lambda$. At smaller angular scales, where lensing
is important,  raising or lowering $\Omega_m h^2$ results in increasing
or decreasing the mass power spectrum, and therefore the  strength of
the  lensing distortion. The question of what can be determined by the
CMB alone becomes academic if we permit unphysically low values of
$\Omega_b$ or $h$, in which case the effect of lensing becomes
negligible. In determining this degeneracy region, then, we have
permitted a conservative, physically-motivated range for the  
cosmological parameters, $h \ge 0.5$ and $\Omega_b h^2 \in [0.016,
0.024]$. This explains the shape of the $\ell_{\rm max}=1500,\,2000$
boundary; for the allowed range of cosmological parameters, the lensing
is strong, breaking the degeneracy when the amplitude of the mass power
spectrum and amplitude of acoustic oscillations are large. Note that the
limiting cases $\Omega_m \to 1,\,0$ are special. In the former case, a
negligible amount of quintessence is present, so that all QCDM models
generate the same, degenerate CMB anisotropy pattern. In the latter case,
as $\Omega_m \to \Omega_b$, the strength of the baryon-photon
oscillations grows, compensating for the decrease in the mass power
spectrum amplitude, so that lensing breaks the degeneracy.

A degeneracy curve represents the  center of a strip of models in the
$\Omega_m - w$ plane which cannot be distinguished by the CMB  alone. To
estimate the width of the degeneracy strip,  we select a quintessence 
and $\Lambda$ model on a given degeneracy curve, vary $\Lambda$,   and 
compute the likelihood that the quintessence model and  the  $\Lambda$
model are distinguishable, allowing for cosmic variance   uncertainty.
For each value of the cosmological constant $\Lambda$, the parameters
$n_s$, $h$, $\Omega_m$ and $\Omega_b$ are varied until the likelihood is
minimized. To compute the likelihood, a novel  estimating procedure has
been  introduced which applies to more general examples of CMB analysis.
The attractive feature is that the likelihood is simple to calculate
analytically, avoiding the need for Monte Carlo.  Suppose Models $A$ and
$B$ are to be compared.  We wish to estimate the likelihood that a Model
$A$ real-sky would be confused as Model $B$. Since the prediction of
Model $A$ is itself non-unique, subject to cosmic variance (and, in
general, experimental error), we need to average the log-likelihood over
the  probability distribution associated with $A$. Only cosmic variance
error,    $C_{\ell}/\sqrt{2 \ell+1}$, is assumed for each multipole
$C_{\ell}$ and  the  distribution is chi-squared. In our notation, 
$C_{\ell}$'s  are the cosmic mean values and  $x_{\ell}$  are the values
measured within our Hubble horizon. Then,  the ``average log-likelihood"
is defined to be  
\begin{equation} {\cal L}_{ba} = \int{\log \frac{{\cal
P}(\{x_{\ell}\} | B)}{{\cal P}(\{x_{\ell}\} | A)} {\cal P}(\{x_\ell\} |A)
dx_1 \ldots dx_{\ell} \ldots } 
\end{equation} 
where ${\cal P}(\{x_\ell\} | A)$ is the probability  of observing the set
of multipoles $\{x_\ell\}$ in a realization of Model $A$. Since each
multipole $C_\ell$ from a full sky map 
is statistically independent,  ${\cal P}(\{x_\ell\} | A)$ can
be written as a simple product of  chi-squared distributions for each
$\ell$. Substituting the  chi-square distribution for ${\em P}(x_\ell |
A)$, ${\cal L}_{ba}$ reduces to
\begin{equation}
{\cal L}_{ba} = - \sum_{\ell } (l+ \frac{1}{2}) \times (1- \frac{C_{\ell a}}{C_{\ell b}} + \log
\frac{C_{\ell a}}{C_{\ell b}}).
\end{equation}
Here we have assumed no experimental error, but it is a simple matter to 
include an additional experimental variance. Note that ${\cal L}_{ba} \ne
{\cal L}_{ab}$ in general, although the difference is small in practice.
We decide distinguishability  according to  the min(${\cal L}_{ba}, \,
{\cal L}_{ab}$). For variations  $\Delta \Omega_{m}$ greater than $\pm
0.05$ from the degeneracy curve value, the  log-likelihood satisfies
$-{\cal L} \ge 6$, corresponding to    distinguishability  at the $3
\sigma$ level or greater. This is the condition we use to determine
distinguishability.

As long as only geometrical effects are important, distinguishability of
a pair of cosmological models entails comparing the shapes of the two
spectra without specifying any normalization. Once lensing and other
non-linear effects are included, the absolute level of anisotropy must be
specified; the shape of the $C_\ell$ spectrum is affected by the absolute
scale of the mass power spectrum. Because the normalization is not known,
some allowance must be made for this uncertainty in the lensing
contribution. For the purposes of this investigation, we have assumed
that for each pair of models, the mass power  spectrum of the first model
is COBE normalized,\cite{BunnWhite} while the spectrum of the second
model must lie within $2\sigma$. A tighter constraint may arise from
future satellite experiments, once the absolute level of anisotropy is
known with better precision. It is interesting to note that despite the
fact that the lensing tends to smear out  sharp features in the CMB
spectrum, effectively  destroying information, we actually gain knowledge
of the underlying mass power spectrum.  As a result, the degeneracy
region shrinks as the effect of lensing accumulates.

Now consider the situation in several years' time, in which the CMB
anisotropy measurements conform closely with one of the degeneracy curves
in Figure 1a, a possibility consistent with current
observations.\cite{WCOS}  The degeneracy means that one cannot
distinguish whether the missing energy is quintessence or vacuum energy.
Furthermore, $\Omega_m$ and $h$ vary along the degeneracy curve (so as to
keep $\Omega_m h^2$ constant), such that the uncertainty in these key
parameters is very large.  How can the ambiguities be resolved?

\begin{figure}
\epsfxsize=3.3 in \epsfbox{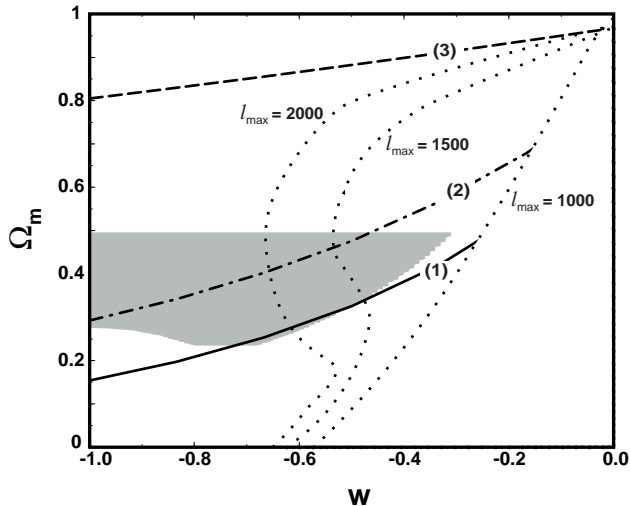}
\caption{The CMB anisotropy constrains models to a particular degeneracy
curve and, independently, provides tight constraints on $n_s$, $\Omega_m
h^2$ and $\Omega_b h^2$.  The latter constraints, along with other
observational limits discussed in the text, fixes an allowed range of
$\Omega_m$ and $w$ (the shaded region using the example discussed in the
text). The combination determines the best-fit models. }
\end{figure}

Other cosmological observations may not be as precise as those of the
CMB  anisotropy, but they have the advantage that they do not share the
same degeneracy. If other observations can be used to determine
separately $\Omega_m$ or $h$ (or some combination of $\Omega_m$ and $h$
other than  $\Omega_m h^2$), then perhaps the degeneracy between
$\Lambda$ and quintessence can be broken. We have considered the current
restrictions on $\Omega_m$ and $h$ obtained by combining the best limits
on age ($> 10$~Gyr), Hubble constant,  baryon fraction ($\Omega_b
h^{3/2}/\Omega_m \sim$~3-10\%), cluster abundance and
evolution,\cite{WangSteinhardt} Lyman-$\alpha$ absorption,\cite{JMEEtAl}
deceleration parameter\cite{supernova} and the mass power spectrum (APM
Survey).\cite{Peac97} The current constraints and the techniques for
combining them have been detailed  elsewhere.\cite{Ostriker,WCOS}   We
also include the fact that the CMB anisotropy  will provide  tight
constraints on $n_s$  and the combinations $\Omega_m h^2$ and $\Omega_b
h^2$  to within a  few
percent.\cite{Sperg97,BondEtAl,JungEtAl,4by4_parameters}  

Even combining all the observational information  listed above,
$\Omega_m$ and $h$ are not highly constrained. Assume for illustrative
purposes that the CMB anisotropy converges on  $n_s=1$, $r=0$, $\Omega_b
h^2 =0.02$ and $\Omega_m h^2 =0.15$ (reasonable values). Then Figure 2 
shows the shaded region in the $\Omega_m$-$w$ plane which can satisfy the
observational   constraints at the $2 \sigma$~level.  In this case, 
acceptable models must lie at the overlap of the  degeneracy curve picked
out by the CMB anisotropy and  the shaded region.

Three possibilities emerge, as shown in Figure 2:  (1) the degeneracy
curve overlaps the  shaded region  only over a limited range of $w$ so
that the ambiguity between quintessence and $\Lambda$ is broken and
$\Omega_m$, $h$ and $w$ are well-constrained; (2) the degeneracy curve
cuts through the  shaded region in such a way that a substantial
ambiguity remains; or (3) the degeneracy curve and the shaded region do
not overlap at all. Case (3) appears at first to be  a contradiction: the
CMB spectrum conforms to the predictions of a  $\Lambda$CDM or QCDM
model, but constraints from other cosmological observations (shaded
region) suggest that the $\Omega_m$ is too small (or too big). However,
this situation is precisely what ought to occur if one of our underlying
assumptions is incorrect: namely, the flatness assumption, constraint
(a). By introducing spatial curvature as an additional component ($A\ne
1$) further degeneracy arises. Associated with curve (3) is a continuous
family of degeneracy curves in the $\Omega_m$-$w$ plane each  beginning
from a different value of $\Omega_m$ along the $w=-1$
axis.\cite{Sperg97,Huey98} Making the universe open (closed) produces
CMB  degeneracy curves beginning  with smaller (larger) values of
$\Omega_m$, whereas  the shaded region in Fig.~2  is only modestly
changed. So, for example, curve (3) in Figure 2 is also degenerate with 
an open model with $\Omega_m=0.4$, $\Omega_{\Lambda}=0.54$ and $h=0.8$,
which is  consistent with the shaded region. Adding curvature is
inconsistent with standard inflation-based models,  but  case (3)
exemplifies how we may be forced observationally to consider the
possibility.

\begin{figure}
\epsfxsize=3.3 in \epsfbox{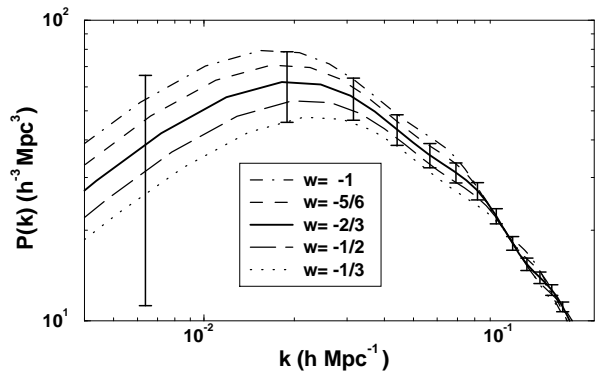}
\caption{Mass power spectra for the models along the 
CMB degeneracy  curve ($\Omega_m h^2 = 0.13$ and $\Omega_b h^2 = 0.02$)
in Fig.~1a are difficult to distinguish with large-scale structure
measurements.  The  error bars are  projected standard errors (1$\sigma$)
for  SDSS assuming the middle ($w=-2/3$) curve. }
\end{figure}

The fact that Case (2) --  continued degeneracy --  remains possible
after  so much data has been invoked is remarkable.  A reduction in
experimental uncertainty ($\sigma$) by a factor of  two for all of the
measurements reduces the size of the shaded region in Fig.~2, but this is
not sufficient to remove all possible degeneracy. For some constraints,
much more than a factor of two improvement  can be anticipated. For
example, the Sloan Digital Sky Survey (SDSS) will provide a substantial
improvement in measurements of the mass power spectrum $P(k)$ and 
velocities,\cite{Vogel95,Teg97} especially on large lengths where $P(k)$
for models along the degeneracy  curve  are most different. Even so, as
Figure 3 shows, the SDSS will not be enough to  resolve the  differences
in the shape of $P(k)$ among models along the degeneracy curve. What
would contribute immensely to the breaking of the degeneracy, rather,  
is an accurate determination of the mass power spectrum on scales  $k
\lesssim 0.1$~h/Mpc, where the models are most different. 

It has been argued that combining the results of future Sloan and CMB
experiments can lead to a substantial improvement in precision, isolating
a narrow region of cosmological parameter space in order to distinguish
between quintessence and Lambda.\cite{4by4_properties,4by4_parameters}  
These improvements rely on the treatment of the errors as statistical
and  independent. We take a more conserative stance that the errors will
be dominantly systematic.  Hence, our conclusions are based on models
passing the tests independently rather than by statistically combining
the CMB and Sloan tests. 

\begin{figure}
\epsfxsize=3.3 in \epsfbox{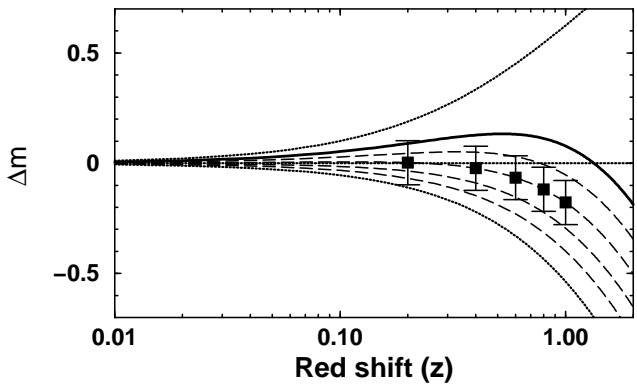}
\caption{  The  magnitude-red shift relation may be a tool for
distinguishing $\Lambda$ models (thick solid curve) from the family of
quintessence models (dashed curves) along the degeneracy curve.  $\Delta
m$ is the difference in the predicted magnitude of a standard candle for
a given model and an open universe ($\Omega_m \rightarrow 0$, middle
dotted curve).  The dashed curves are QCDM models with $w=-5/6, -2/3,
-1/2, -1/3$ from top to bottom, respectively. Hypothetical type IA
supernova data are shown at several red shift, assuming $w=-2/3$ with
$1\sigma$ error bars of $\pm 0.1$ magnitudes. For reference, an
$\Omega_{\Lambda}=1$ (upper dotted) and  $\Omega_m=1$ (lower dotted) flat
model are shown. }
\end{figure}

Figure 4 shows the  prediction for the red shift luminosity relation,
measured using Type IA supernovae as standard candles\cite{supernova} for
the same models along the degeneracy curve. In this case the 
quintessence models are more distinct from the $\Lambda$ model; however,
it is premature to  say whether observations will become accurate enough
to make this measurable. Not only will a large number of high red shift
SNe have to be observed, but the systematic errors in the magnitude
calibration will have to be reduced, to $\Delta m \lesssim 0.1$, in order
that a turn-over in $\Delta m$ is well determined.

One might expect that ground-based CMB experiments, which probe smaller
angular scales than accessible by satellite experiments, can dramatically
resolve the degeneracy problem. It is precisely on the small angular
scales that non-linear effects such as gravitational lens distortion, 
the Rees-Sciama\cite{ReesSciama}  and 
Ostriker-Vishniac\cite{OstrikerVishniac} effects  are important. However,
these effects depend not only on the broad cosmological parameters
$\Omega_m, \, \Omega_b,\, h$, but also on the details of re-ionization
and small scale structure formation, about which there probably remains
enough uncertainty to prevent this method from being used as a fine model
discriminant. It is not clear whether such constraints, while sufficient
to differentiate between $\Lambda$CDM and SCDM, can discriminate between
quintessence and $\Lambda$. 

Our conclusion is asymmetrical. A large class of quintessence models,
those with rapidly varying $w$ or constant $w \gtrsim - \Omega_Q/2$,  can
be   distinguished from $\Lambda$ models by near future CMB experiments
such as MAP.  However,  any given $\Lambda$ model   is indistinguishable
from  the subset of quintessence models along its degeneracy curve. CMB
experiments which probe small angular scales where gravitational lens
distortion is expected to be important, such as Planck, can be expected
to cut into the degeneracy region. Combining the constraints which  the
CMB imposes on $n_s$, $\Omega_m h^2$ and $\Omega_b h^2$ to the other
current observational constraints sometimes, but not always, breaks the
degeneracy. Adding spatial curvature as an additional degree of freedom
increases the degeneracy. Depending on how measurements overlap, new
observational  techniques must be invented to break the degeneracy.

We thank J. Bahcall, Wayne Hu, W. Press, J.P. Ostriker and M. Strauss  
for many  useful comments.  We thank M. Vogeley for explaining
anticipated errors in SDSS and  providing a code to estimate their
magnitude. This research was supported by the Department of Energy at
Penn, DE-FG02-95ER40893. We have modified the CMBFAST software
routines\cite{CMBFAST} for our numerical computations.

%%%%%%%%%%%%%%%%%%%%%%%%%%%%%%%%%%%%%%%%%%%%%%%%%%%%%%%%%%%%%%%%%%%%%%%%%%
%%%%%%%%%%%%%%%%%%%%%%%%%%%%%%%%%%%%%%%%%%%%%%%%%%%%%%%%%%%%%%%%%%%%%%%%%%

%%%%%%%%%%%%%%%%%%%%%%%%%%%%%%%%%%%%%%%%%%%%%%%%%%%%%%%%%%%%%%%%%%%%%%%%%%

\vfill

%%%%%%%%%%%%%%%%%%%%%%%%%%%%%%%%%%%%%%%%%%%%%%%%%%%%%%%%%%%%%%%%%%%%%%%%%%
\end{document}